\documentclass{article}

\usepackage{psfrag,latexsym,amssymb,amsmath,color,epstopdf,lineno}
\usepackage{graphicx}

\usepackage{amsthm}
\newcommand{\vX}{\boldsymbol{X}} 
 
\newcommand{\bQ}{\mbox{\boldmath $\mathsf{Q}$}}
\newcommand{\bq}{\boldsymbol{q}}

\def\bra{\langle}
\def\ket{\rangle}
\def\p{\partial}

\def\beq{\begin{equation}}
\def\eeq{\end{equation}}
\def\la{\label}

\def\r#1{(\ref{#1})}
\newcommand{\mylab}[3]{\raisebox{#2}[0mm][0mm]{%
\makebox[0mm][l]{\hspace*{#1}{#3}}}}%
\def\utau{u_\tau}

\def\figpath{./}
\def\spacce#1{\hskip #1pt}
\def\drawline#1#2{\raise 2.5pt\vbox{\hrule width #1pt height #2pt}}
\def\solid{\drawline{24}{.5}\nobreak}

\def\bdash{\hbox{\drawline{5}{.5}\spacce{2}}}

\def\dashed{\bdash\bdash\bdash\nobreak}

\def\bdot{\hbox{\drawline{1}{.5}\spacce{2}}}

\def\dotted{\hbox{\leaders\bdot\hskip 24pt}\nobreak}

\def\chndot{\hbox%
{\drawline{4.6}{.5}\spacce{2}\drawline{1}{.5}\spacce{2}\drawline{4.6}{.5}\spacce{2}\drawline{1}{.5}\spacce{2}\drawline{4.6}{.5}}\nobreak }
\def\circle{$\circ$\nobreak }

\def\trian{\raise 1.25pt\hbox{$\scriptstyle\triangle$}\nobreak}

\def\dtrian{\raise 1.25pt\hbox%
{$\scriptscriptstyle\bigtriangledown$}\nobreak}
\def\rtrian{\raise 1.25pt\hbox%
{$\scriptscriptstyle\vartriangleright$}\nobreak}

\def\squar{\raise 1.25pt\hbox{$\scriptstyle\Box$}\nobreak}

\def\diamon{\raise 1.25pt\hbox{$\scriptstyle\diamond$}\nobreak}

\newcommand{\soliddtrian}{$\blacktriangledown$\nobreak}

\def\linedtri1{\hbox{\bdash\hspace{-1.6mm}$\bigtriangleup$\hspace{-0.8mm}\bdash}\nobreak}
\def\soliddtrian1{$\blacktriangledown$\nobreak}
\def\solidrtrian2{$\blacktriangleright$\nobreak}
\def\solidltrian3{$\blacktriangleleft$\nobreak}

\def\citep#1{\cite{#1}}
\title{Fake turbulence}

\author{Javier Jim\'enez\\
School of Aeronautics, U. Polit\'ecnica Madrid, 28040 Madrid Spain}
\date{\today}

\begin{document}
\maketitle
\begin{abstract}
High-dimensional dynamical systems projected onto a reduced-order model cease to be
deterministic and are best described by probability distributions in state space. Their
equations of motion  map onto an evolution operator with a deterministic component describing the projected
dynamics, and a stochastic one from the neglected
dimensions. It is shown that, for projections in which the deterministic component is dominant,
`physics-free' stochastic Markovian models can be constructed that mimic many of the statistics of
the real flow, even for fairly crude operator approximations. Deterministic models converge
to steady states. This is related to general properties of Markov chains and illustrated with
data-driven models for a moderate-Reynolds number turbulent channel.
\end{abstract}


\section{Introduction}\la{sec:intro}

The recent proliferation of language models that mimic human conversation based on
the statistical analysis of largely syntax-free unlabelled data \citep{GPT3:2020},
naturally raises the question of whether something similar can be done in physics. In both cases,
the underlying dynamics is known: grammar in the former and the equations of motion in the latter,
and the goal is not so much to reproduce the system in detail as to construct statistical models
that either simplify simulations or isolate aspects of the problem to be further studied by
other means. Rather than attacking the abstract question, we proceed to construct
such a model in the restricted domain of wall-bounded turbulence; a physical system for which the
equations are understood, and where the problem is how to interpret the observations from
numerical simulations and experiments. A basic statistical analysis of such a system was
performed in \cite{jim23:causflu}, and will not be repeated here. Our goal is to construct simple
models in a controlled environment in which the right answers are essentially known, and
learn what can be reproduced from data statistics, which simplifications are admissible, and
what errors are introduced by them. On the way, we will learn something about what data-driven
models can do in general.

Turbulence is a dynamical system whose temporal evolution follows the deterministic Navier--Stokes
equations. However, when the state of the flow, $\vX$, is parametrized, for example, by the velocity
components at all points of the flow field, its dimensionality is formally infinite and, even when
numerically discretized, the number of degrees of freedom is typically in the millions (see table
\ref{tab:cases}). Its direct numerical simulation (DNS) is a well-developed technique, but
high-dimensional systems are generically chaotic, and the trajectories thus obtained are only
relevant in the sense of being statistically representative of many possible such trajectories.
Interpreting DNS data usually implies projecting them onto a lower-dimensional manifold of
observables, whose evolution is no longer deterministic because each projected point represents many
different states along the neglected dimensions. It was argued in \cite{jim23:causflu} that these
reduced-order models (ROMs) are best studied by replacing the equations of motion with transition
probabilities between ensembles of projected states at different times.

This statistical view of turbulence physics has a long history, although computers have only
recently been able to deal with the required data sets. Early work treated the flow either as a
large but finite collection of coherent structures \citep{onsag49}, or as the evolution of ensembles
in functional space \citep{hopf52}. More recent analyses have centred on the probability
distributions over finite-dimensional partitions of the state space, for which the evolution reduces
to a map between temporally equispaced snapshots. However, while the classical statistical analysis
applies to dynamical systems in which the probabilistic description is a choice, we will be
interested in the intrinsically non-deterministic case of ROMs. Related operator methods in
turbulence are discussed by
\cite{KaiserNoack_JFM14,schmid:iop18,Brunt:ARFM20,FernexNoack:21,Taira22} and \cite{Souza23}, among
others. This paper explores how much of the behaviour of a turbulent system can be approximated by a
time series generated by the transition operator that links consecutive time steps in a `training'
experiment. In essence, which part of the long-term behaviour of the system is contained in its
otherwise `physics-free' short-term statistics.

The paper is organised as follows. The methods and data used in the analysis are described in
\S\ref{sec:methods}. Results are discussed in \S\ref{sec:results}, including how well physics is
approximated by the model time series, the reasons why it does or does not, and what measures can be
taken to alleviate hallucinations and overfitting. Conclusions are summarised in \S\ref{sec:conc}.
  
\begin{table}
  \begin{center}
    \def~{\hphantom{0}}
    \begin{tabular}{lcccccl}
      Case  &      $Re_\tau$ & $u_\tau\tau/h$  & $n_T$ & Grid  & Deg. of freedom \\
\hline\\[-3pt]%
      C350          & 350& 0.022 & $2.5\times 10^4$ & $64 \times 193 \times 64$   & $1.6\times 10^6$ \\
      C550          & 535 & 0.020 & $2.5\times 10^4$ & $96 \times 257 \times 96$   & $4.7\times 10^6$\\
      C950          & 949 & 0.025 & $5.1\times 10^4$ & $128\times 385\times 128$   & $11.7\times 10^6$\\
  \end{tabular}
  \end{center}
\caption{Parameters of the DNS data bases. The number of flow snapshots is $n_T$, spaced in time by
$\tau$. Since the two walls are treated as independent, the effective number of data
points is $2 n_T$. The grid is expressed in terms of real Fourier or Chebychev $(x,y,z)$ modes, and
the number of degrees of freedom is twice the number of grid points.
 }%
\label{tab:cases}
\end{table}

\section{Methods and data}\la{sec:methods}

We analyse an extended version of the computational data set in \cite{jim23:causflu}. The number of
snapshots in that simulation (C950, see table \ref{tab:cases}) has been extended to improve
statistics, and two simulations (C350 and C550) have been added to explore the effect of the
Reynolds number. A pressure-driven spatially periodic turbulent channel flow is established between
parallel plates separated by $2h$. The streamwise, wall-normal, and spanwise coordinates are $x, y$
and $z$, respectively, and the corresponding velocity components are $u, v$ and $w$. Capital letters
denote $y$-dependent ensemble averages, $\bra\ket$, and lower-case ones are fluctuations with
respect to them. Primes are root-mean-squared intensities, and the `+' superscript
denotes normalisation with the kinematic viscosity, $\nu$, and with the friction velocity $\utau = \sqrt{\nu
\p_y U|_{y=0}}$. The code is standard dealiased Fourier--Chebychev spectral \citep{kmm}, with
constant mass flux. Time expressed in eddy turnovers is denoted as $t^*=\utau t/h$, and the
friction Reynolds number is $Re_\tau = h \utau/\nu$. Further details can be found in
\cite{jim13_lin}. We mostly describe results for the highest-Reynolds number case C950, but
they vary little within the limited range of $Re_\tau$ available, and some comparisons are included.

The wall-parallel periods of the computational box, $L_x = \pi h/2$ and $L_z = \pi h/4$, are chosen
small enough for the flow to be minimal in a band of wall distances $y/h\approx 0.2 -0.6$
\citep{oscar10_log}, in the sense that a non-negligible fraction of the kinetic energy is contained
in a single large structure that bursts irregularly at typical intervals $t^* \approx 2$--3. The
present simulations contain several hundreds of bursts per wall, and about 100 samples per burst.
Moreover, since the small box allows little interaction between the two walls, they are treated as
independent, doubling the number of effective snapshots. The numerical time step, 80 to 500 times
shorter than the time between snapshots, depending on $Re_\tau$, varies by about $\pm 15\%$ in the
run, and data are interpolated to a uniform interval $\tau$ before their use.

We compile statistics over partitions in which each coordinate of the $D$-dimensional state space is
discretised in $O(m)$ bins, so that the dimension of the probability distribution, $\bq=\{ q_j\},\,
j=1\ldots N_D$, is $N_D=O(m^D)$. Snapshots separated in time by $\Delta t$ are related by a
Perron-Frobenius transition operator, $\bQ_{\Delta t}$ (PFO), see \citep{BeckSchl93}, defined by
\beq
\bq(t+\Delta t) = \bQ_{\Delta t} \bq(t),
\la{eq:markov0}
\eeq
which is an $N_D\times N_D$ stochastic matrix with $O(m^{2D})$ elements. Each of its columns is the
probability distribution of the descendants at $t+\Delta t$ of the points in one cell at time $t$
\citep{lancaster}. We will assume $\bQ$ to depend on the data interval $\Delta t$, typically chosen
as a multiple of the sampling interval, but not explicitly on time. It is constructed as the
joint histogram of the indices of the cells occupied by snapshots separated by $\Delta t$ during a
training run \citep{Ulam1964}. Since the number of data needed to populate $\bQ$ is at
least a multiple of the number of its elements, even a modest choice, $m\sim 10$, limits the
dimension of a model trained on $10^5$ data to $D\leq 3$. We use $D=2$ in the paper.

The first task is to reduce the dimensionality of the space from the DNS resolution to the much
smaller ROM without losing too much dynamics. This was done in \cite{jim23:causflu} by a combination
of physical reasoning and computational testing. Briefly, we retain nine Fourier modes along the $x$
and $z$ directions, integrated over the band of $y$ in which the flow is minimal. The result is a
list of modal amplitudes, $I_{* ij}$, and inclinations, $\psi_{* ij}$, where the subindices refer to
the velocity component $(*)$, and the Fourier modes $(i,j)$ involved. The reader is referred to
\cite{jim23:causflu} for a more detailed discussion. We simply treat them here as 44 physically
interpretable summary variables that describe the evolution of bursting in our data base
\citep{orr07a,jim13_lin,encinar:20}.

\begin{figure}
\centering
\raisebox{0mm}{\includegraphics[height=.30\textwidth,clip]%
{\figpath 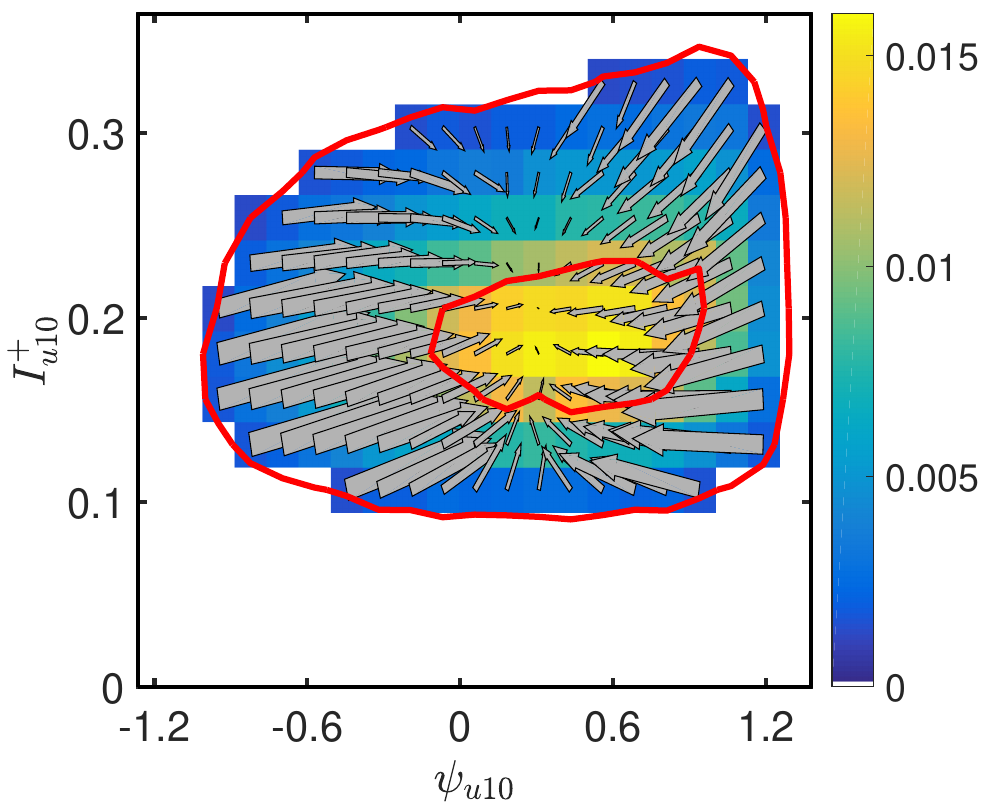}}%
\mylab{-.29\textwidth}{.26\textwidth}{(a)}%
\hspace*{1mm}%
\raisebox{0mm}{\includegraphics[height=.30\textwidth,clip]%
{\figpath 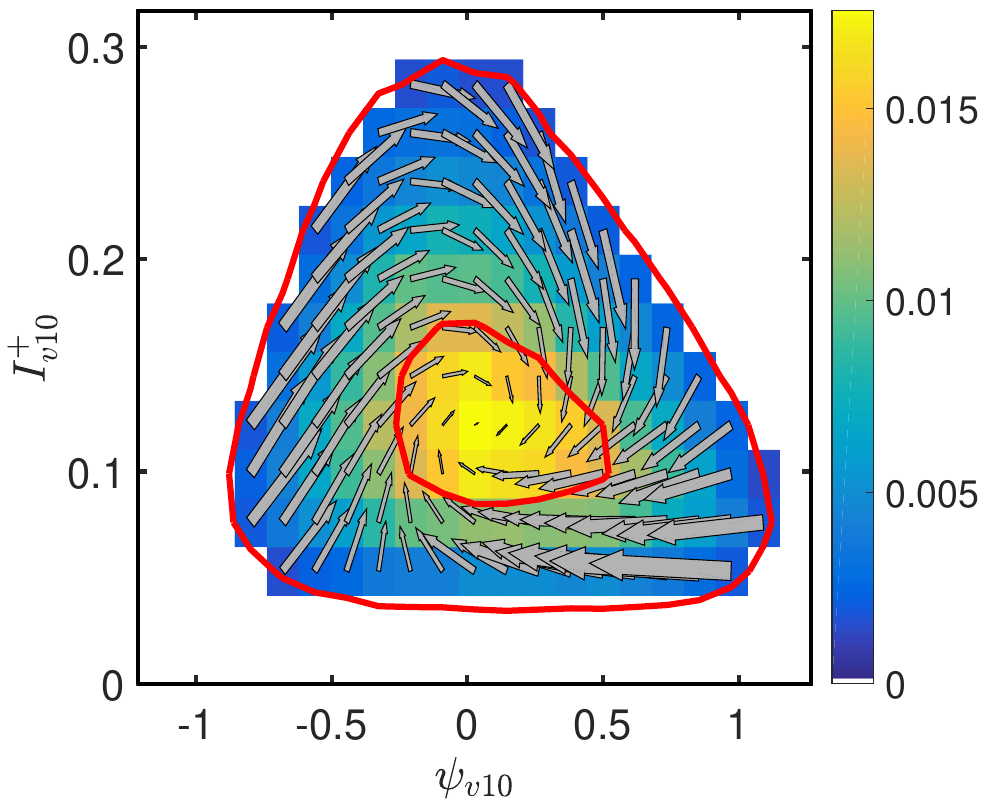}}%
\mylab{-.29\textwidth}{.26\textwidth}{(b)}%
\caption{%
Probability distribution and flux vectors for two pairs of projection variables. The arrows link 
the centre of each partition cell with the mean location of  the flow after a time iteration. The
red contours enclose 0.3 and 0.95 of the probability mass.
(a) A typical disorganised variable pair (Case I). 
(b) The well-organised Orr burst described in the text (Case II).
$\Delta t^*=0.075$. $21\times 20$ cells. C950  \citep[adapted from][]{jim23:causflu}.
}
\label{fig:maps}
\end{figure}

Choosing the best variable pair involves testing 946 possible combinations of two variables. Two
examples of the raw material for these tests are shown in figure \ref{fig:maps}. In each case, the
coloured background is the long-time joint probability distribution of the two variables, compiled
over a regular partition of either $m_1\times m_2=15\times 13$ or $21\times 20$ cells along the
first and second variable. Results are relatively insensitive to this choice. The finer partition
increases the resolution of the results, but decreases their statistical convergence and, to avoid
noisy low-probability cells, we always restrict ourselves to the interior of the probability contour
containing 95\% of the data. Each cell along the edge of this region approximately contains 300 data
points for the coarser partition and 150 for the finer one. The temporal probability flux is
represented by vectors joining the centre of each cell with the average location of the points in
the cell after one time step, and how much dynamics is left in the projected plane can be estimated
from how organised these vectors are \citep{jim23:causflu}. Most cases are as in figure
\ref{fig:maps}(a), where the state migrates towards the high-probability core of the distribution,
essentially driven by entropy. We will denote this variable combination as Case I. A few variable
pairs are more organised, as Case II in figure \ref{fig:maps}(b), whose upper edge is the burst
described above. The inclination angle in the abscissae evolves from a backward to a forward tilt
while the intensity in the ordinates first grows and then decays.

The probability maps in figure \ref{fig:maps} include a deterministic component and a disorganised
one that represents the effect of the discarded variables. The latter typically increases with the
time increment and dominates for $\Delta t^*\ge 0.2$, but we will see below that short time
increments have problems linked with resolution, and that the limit $\Delta t\to 0$ implies
infinitesimally small partition cells. The latter are limited by the amount of training data, and our
models are necessarily discrete both in state space and in time. 

\section{Results}\la{sec:results}

We now describes Markovian models that approximate the order in which the flow visits the partition
cells by iterating \r{eq:markov0}. None of them can fully represent the coarse-grained dynamical
system, which is generally not Markovian \citep{BeckSchl93}, but we will be interested in three
questions. The first is whether the Markov chain converges in the projected subspace to a
probability distribution similar to that of the original system. The second is whether the
projection conserves enough information of the neglected dimensions to say something about them. The
third is whether $\bQ$ can be approximated without destroying its usefulness.
 
\begin{figure}
\centering
\raisebox{0mm}{\includegraphics[height=.30\textwidth,clip]%
{\figpath 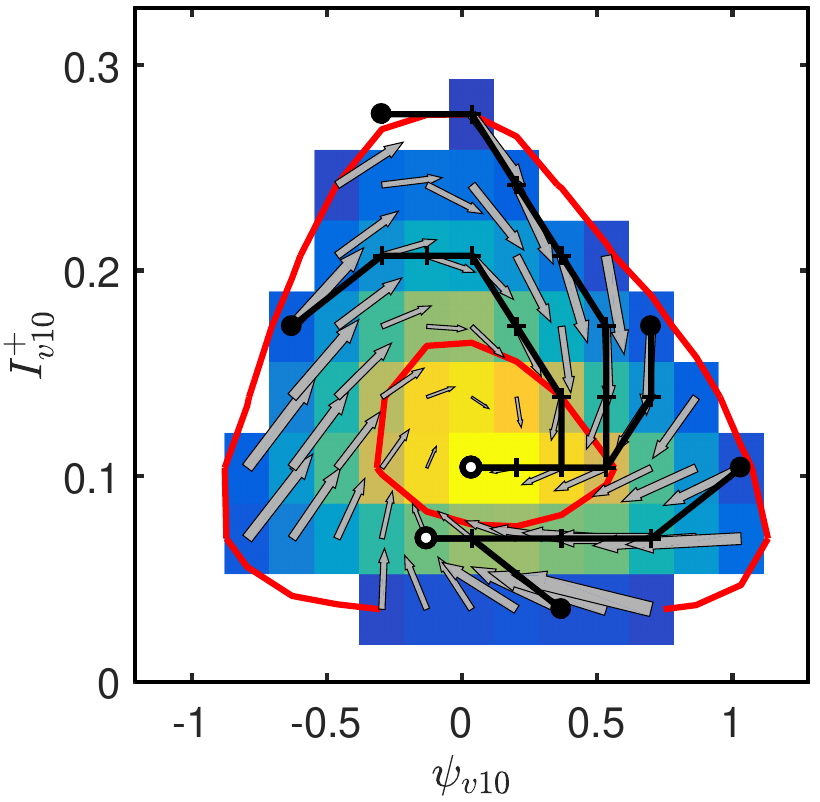}}%
\mylab{-.24\textwidth}{.26\textwidth}{(a)}%
\hspace*{2mm}%
\raisebox{0mm}{\includegraphics[height=.30\textwidth,clip]%
{\figpath 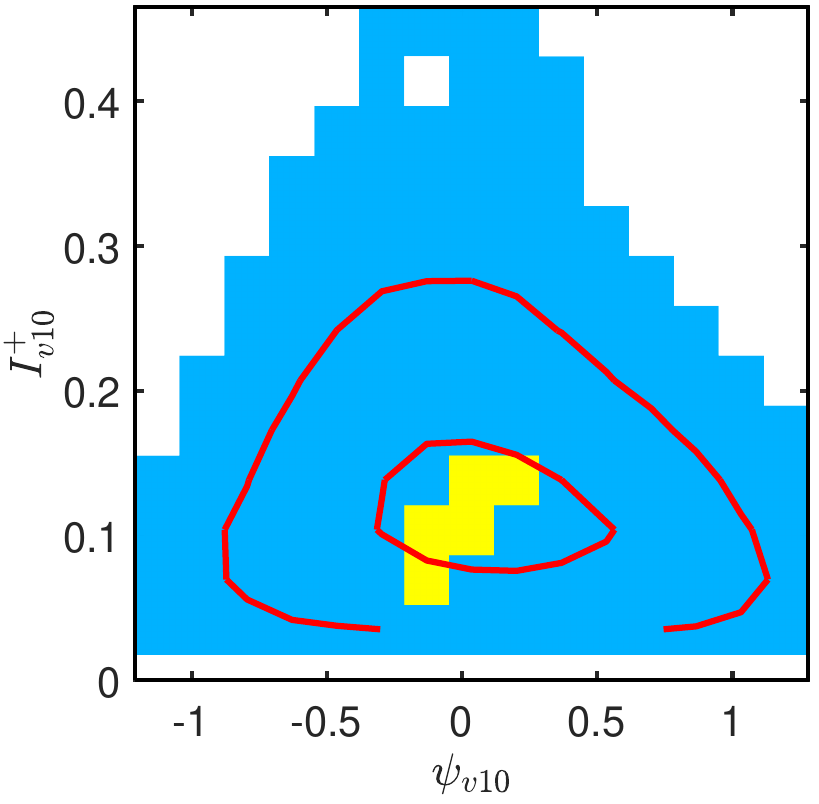}}%
\mylab{-.24\textwidth}{.26\textwidth}{(b)}%
\hspace*{2mm}%
\raisebox{0mm}{\includegraphics[height=.305\textwidth,clip]%
{\figpath 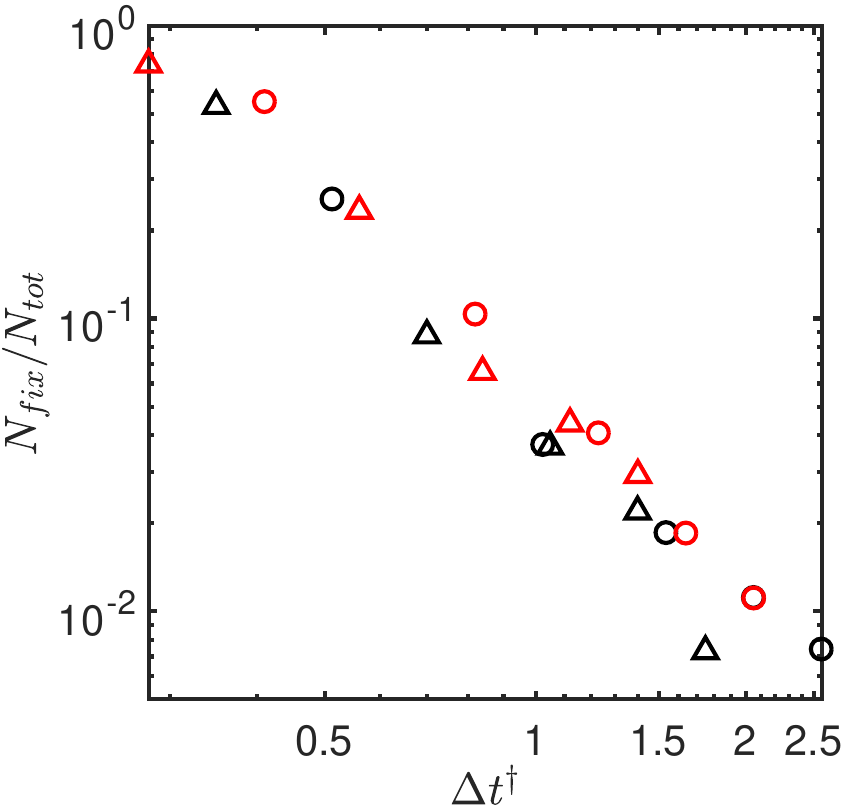}}%
\mylab{-.06\textwidth}{.26\textwidth}{(c)}%
\caption{%
(a) A deterministic reduced-order model for the Case II variables. Several
initial conditions are marked by solid symbols, and the model transitions from a cell at $t$
to the mean expected position of the system at $t+\Delta t$. After some iterations, all trajectories
settle to the cells marked by open symbols. $\Delta t^*=0.075$. C950. $15\times 13$
cells.
(b) Cell classification for the model in (a). White cells are not visited during training. Yellow
cells are absorbers. Blue are regular cells.
(c) Fraction of absorbing cells for different deterministic models. \circle, $21\times 20$ cells; \trian, $15\times 13$. Black, C950; red, C550. 
}
\label{fig:fake1}
\end{figure}

The simplest model based on the transition operator is to substitute time stepping 
by the transition from each partition cell at time $t$ to the cell containing the average position of
its descendants at $t+\Delta t$. Figure \ref{fig:fake1}(a) shows that this is not 
effective. Although the model follows at first the trend of the probability flow, it 
drifts towards the dense core of the probability distribution. Some of the core cells are
absorbers, i.e. they can be entered but not exited, and the model eventually settles into
one of them. Substituting the average position by other deterministic rule, such as the most
probable location, leads to similar results.

General theory requires that, if a model is to approximate the statistics of its training run, it
should not contain absorbing cells \citep{feller1XV}. This depends on the ratio between the time step
and the coarseness of the partition. Intuitively, if the `state-space velocity' of a model is $V_X$
and the `cell dimension' is $\Delta X$, any process with $\Delta t< \Delta X/V_X$ never leaves its
cell.

If we assume that the model explores the $m_p$ cells along the diameter of a partition with a
characteristic time $T_{s}$, the relevant normalisation of the ratio between temporal and spatial
resolution is $m_p\Delta t /T_{s}$. Figure \ref{fig:fake1}(b) shows the cell classification for the
model in figure \ref{fig:fake1}(a). The four yellow cells contain the mean position of their next
iteration, and are absorbers. Figure \ref{fig:fake1}(c) shows the fraction of active cells (i.e.,
those visited during training) that are absorbers, for different Reynolds numbers and partition
resolutions. A dimensionless time based on the resolution along the two coordinate axes, $\Delta
t^\dag = \Delta t^* \sqrt{m_1 m_2}$, collapses the data reasonably well and the figure shows that
the model in figure \ref{fig:fake1}(a) contains at least one absorbing cell even when $\Delta t\gg
\tau$, when there is essentially no dynamics left in the operator. The distance between markers
along the trajectories in figure \ref{fig:fake1}(a) $(\Delta t^\dag \approx 1.5)$ shows how far the
system moves in one modelling step.

\begin{figure}
\centering
\raisebox{0mm}{\includegraphics[height=.30\textwidth,clip]%
{\figpath 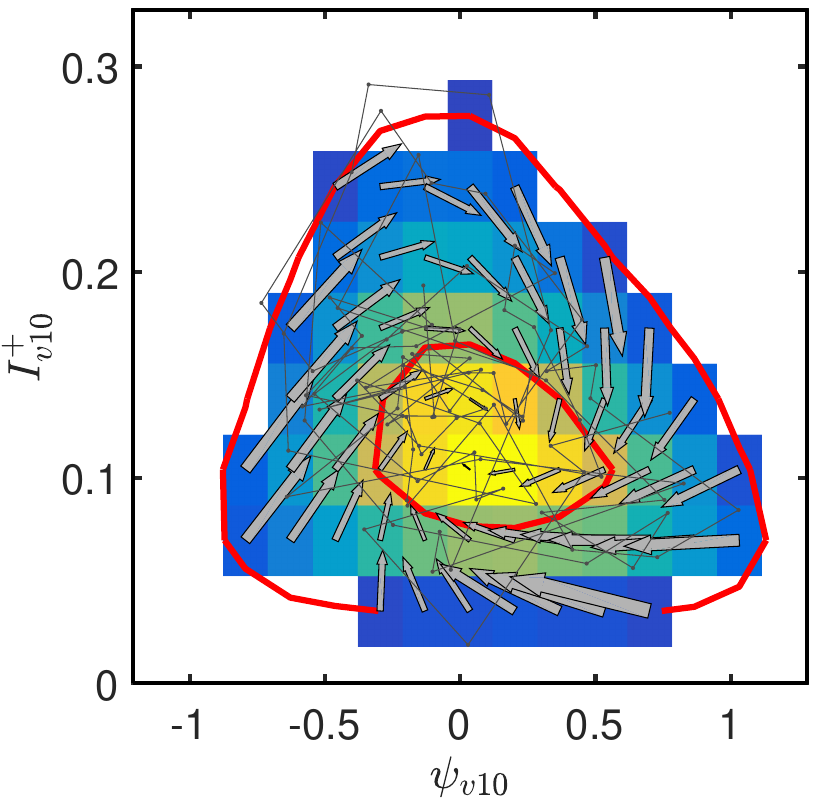}}%
\mylab{-.24\textwidth}{.26\textwidth}{(a)}%
\hspace*{2mm}%
\raisebox{0mm}{\includegraphics[height=.30\textwidth,clip]%
{\figpath 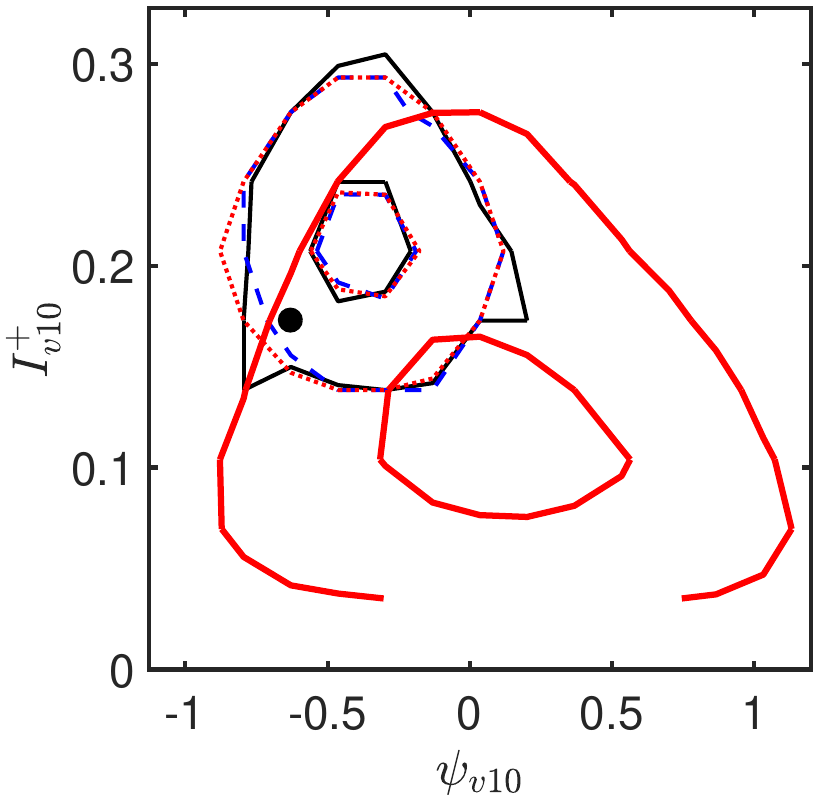}}%
\mylab{-.24\textwidth}{.26\textwidth}{(b)}%
\hspace*{2mm}%
\raisebox{0mm}{\includegraphics[height=.30\textwidth,clip]%
{\figpath 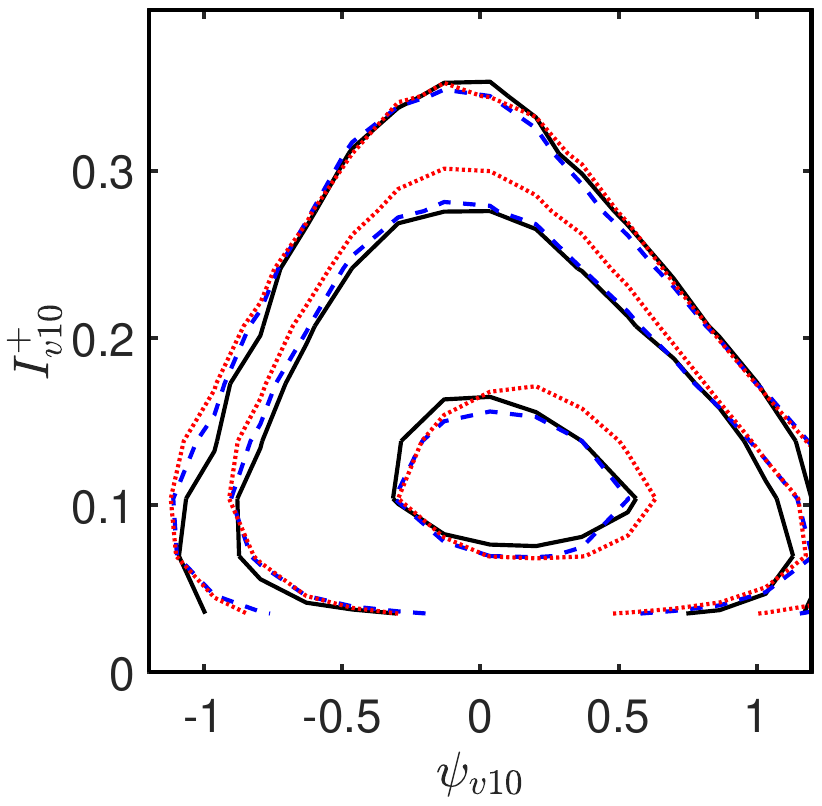}}%
\mylab{-.24\textwidth}{.26\textwidth}{(c)}%
\caption{%
(a) As in figure \ref{fig:fake1}(a), for $n_T=100$ time steps evolved using for each iteration
a randomly selected cell from the PFO probability cloud.
(b) Typical probability distribution of the one-step iterations of the cell marked by a solid circle.
\solid, full PFO model; \dashed, using a Gaussian approximation to the true PFO; \dotted, Gaussian
approximation with fitted parameters. $n_T=10^5$. Contours contain 0.3, 0.95 of the 
probability mass.
(c) Invariant probability densities for the three models in (b). Contours contain 0.3, 0.95, and
0.995 of the probability mass.
In the three figures, $\Delta t^*=0.075$. $15\times 13$ cells. C950.
}
\label{fig:fake_PPF}
\end{figure}

Ergodicity can be restored by including the full probability distribution of the iterates in the
transition operator \r{eq:markov0} instead of deterministic values. The path in figure
\ref{fig:fake_PPF}(a) is a random walk over the cell indices, created by choosing at $t+\Delta t$ a
random cell from the probability distribution in the column of $\bQ_{\Delta t}$ that contains the
descendants of the cell at $t$. In addition, and mostly for cosmetic purposes, the cell selected as
$\bq(t+\Delta t)$ is mapped to a random state within it, $\vX(t+\Delta t)$. As the path explores
state space, it creates a one-step probability map that mimics $\bQ_{\Delta t}$, and counteracts the
entropic drive towards the core of the distribution by adding temperature. The Perron--Frobenius
theorem \citep{lancaster,feller1XV} guarantees that the one-step transition operator determines the
invariant probability density (IPD) of the Markov chain. Under mild conditions that essentially
require that the attractor cannot be separated into unconnected subsets, stochastic matrices have a
unique dominant right eigenvector that can be scaled to a probability distribution, with unit
eigenvalue. Any initial set of cells from non-zero columns converges to this distribution at long
times. Each iteration scheme creates its own PFO and IPD. The invariant density of the deterministic
model in figure \ref{fig:fake1}(a) is the set of absorbing cells, which attract all the initial
conditions. The long-term distribution of the `pre-trained' (PPF) chain in figure
\ref{fig:fake_PPF}(a) is indistinguishable from the data used to train it.

Even if the PPF model is a good representation of the flow statistics, the full transition operator
is a large matrix that has to be compiled anew for each set of flow parameters. Moreover, figure
\ref{fig:fake_PPF}(b) shows that, although the full operator is a complex structure, at least some
of the conditional transition probabilities can be approximated by simpler distributions. The black
contour in this figure is the true distribution of the one-step iterations from the cell marked by a
solid symbol. The dashed contours are a Gaussian approximation to that probability, with the same
first- and second-order moments. The parameters of the Gaussian are smooth functions of the
projection variables, at least within the 95\% core of the IPD, and the dotted contours are also
Gaussian, but using parameters that have been fitted over the whole IPD with a
second-order least-square fit. The three approximations are very similar, and figure
\ref{fig:fake_PPF}(c) shows that their IPDs also agree to fairly low probability levels. We will
mostly present results for the PPF from now on, although keeping in mind that the simpler
approximations may be useful in some cases.


\subsection{Fake physics}\la{sec:fake}

\begin{figure}
\centerline{%
\raisebox{0mm}{\includegraphics[height=.29\textwidth,clip]%
{\figpath 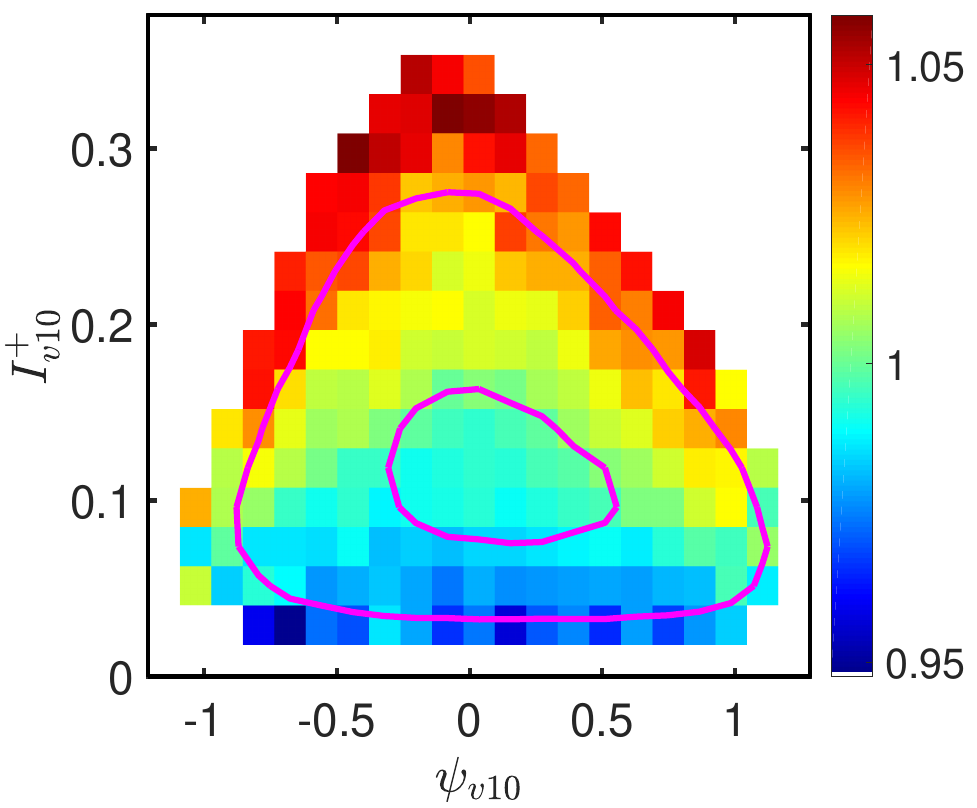}}%
\mylab{-.10\textwidth}{.26\textwidth}{(a)}%
%
\raisebox{0mm}{\includegraphics[height=.29\textwidth,clip]%
{\figpath 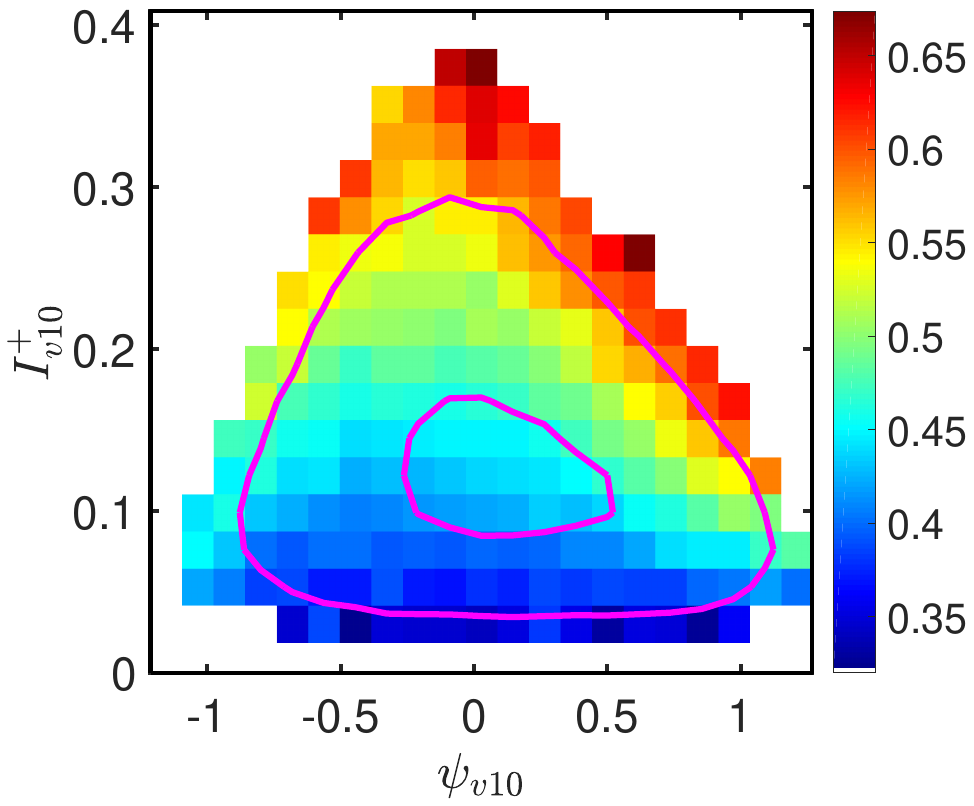}}%
\mylab{-.10\textwidth}{.26\textwidth}{(b)}%
\hspace*{1mm}%
\raisebox{0mm}{\includegraphics[height=.29\textwidth,clip]%
{\figpath 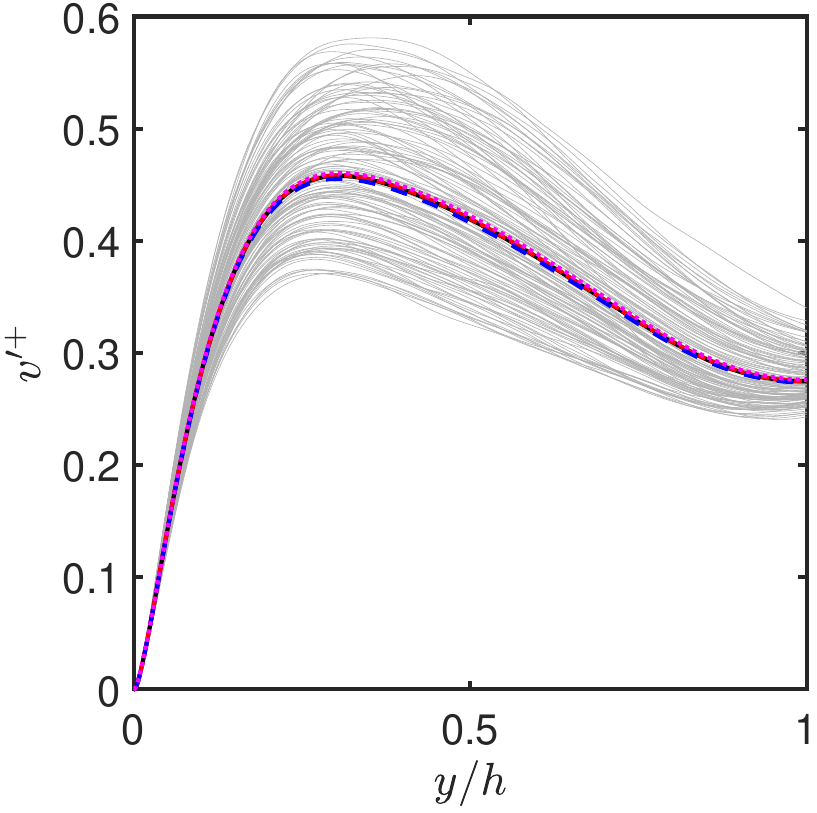}}%
\mylab{-.06\textwidth}{.26\textwidth}{(c)}%
}%
\caption{%
(a) Distribution of the velocity gradient at the wall, $\p_y U$, conditioned to individual cells. 
(b) As in (a), for the maximum of $v'$ of the retained Fourier modes. 
(c) The thicker lines are fluctuation profiles of the wall-normal velocity for the Markovian models
in figure \ref{fig:fake_PPF}, and for the training data. Light grey lines are mean profiles compiled
over individual cells of the two-dimensional invariant distribution. $n_T=10^5$.
$\Delta t^*=0.075$. $21\times 20$ cells. Case II of C950.%
}
\label{fig:fake_prof}
\end{figure}

Figure \ref{fig:fake_PPF}(c) should not be interpreted to mean that the Markovian trajectories are
the same as in turbulence. All models quickly diverge from their training trajectory, and, even if
this is also true for turbulence trajectories starting from the same cell, the model and turbulence
trajectories do not shadow each other.
However, the agreement of the probability densities in figure \ref{fig:fake_PPF}(c) suggest that
some statistical properties of turbulence may be well predicted by the models. This is true for most
of the mean velocity and fluctuation profiles, for which it is hard to distinguish the models from
the data. In some cases, such as the mean velocity and the intensity of the fluctuations of the
wall-parallel velocity components, this is because the flow statistics are relatively insensitive to
the position in the projected subspace. An example is the distribution of the wall shear in figure
\ref{fig:fake_prof}(a). In others, such as the wall-normal velocity fluctuation intensities in
figure \ref{fig:fake_prof}(b), the agreement depends on the convergence of the probability density.
Note the different range of the colour bars in figures \ref{fig:fake_prof}(a) and
\ref{fig:fake_prof}(b). Figure \ref{fig:fake_prof}(c) shows that even in the case of $v'$, the
fluctuation profile is well represented by the stochastic models. The light grey lines in this
figure are intensity profiles for flows that project onto individual cells of the IPD. The darker
lines, which are compiled for the training data and for the three stochastic Markov models, are
long-time averages. They are indistinguishable from each other, even if the profiles belonging to
individual cells are quite scattered, and the mean values only agree because the Markov chain
converges to the correct probability distribution.

\begin{figure}
\centerline{%
\raisebox{0mm}{\includegraphics[height=.30\textwidth,clip]%
{\figpath 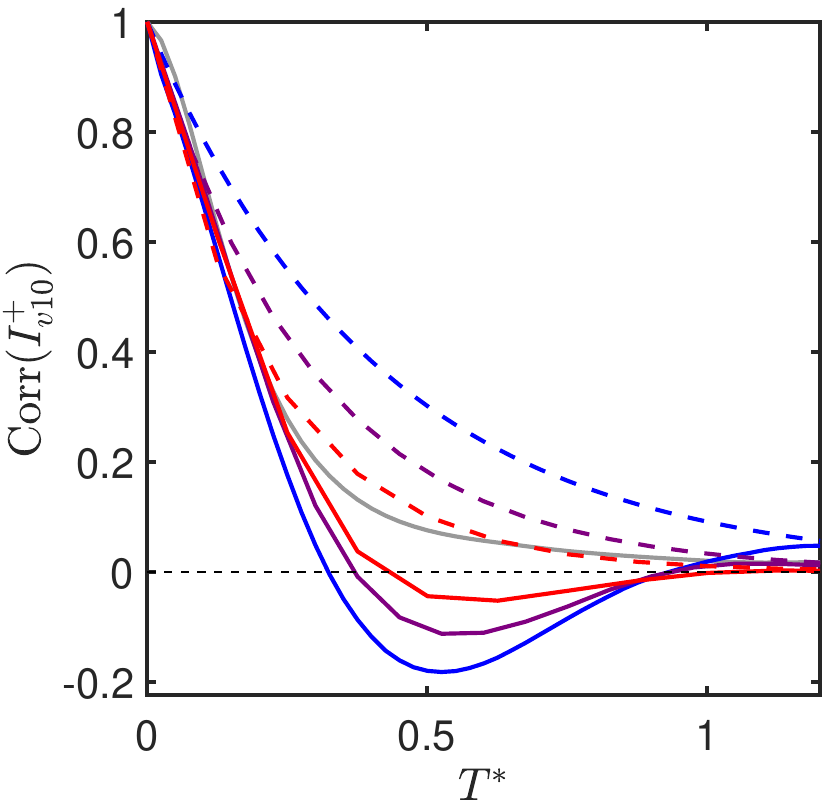}}%
\mylab{-.04\textwidth}{.26\textwidth}{(a)}%
\hspace*{2mm}%
\raisebox{0mm}{\includegraphics[height=.30\textwidth,clip]%
{\figpath 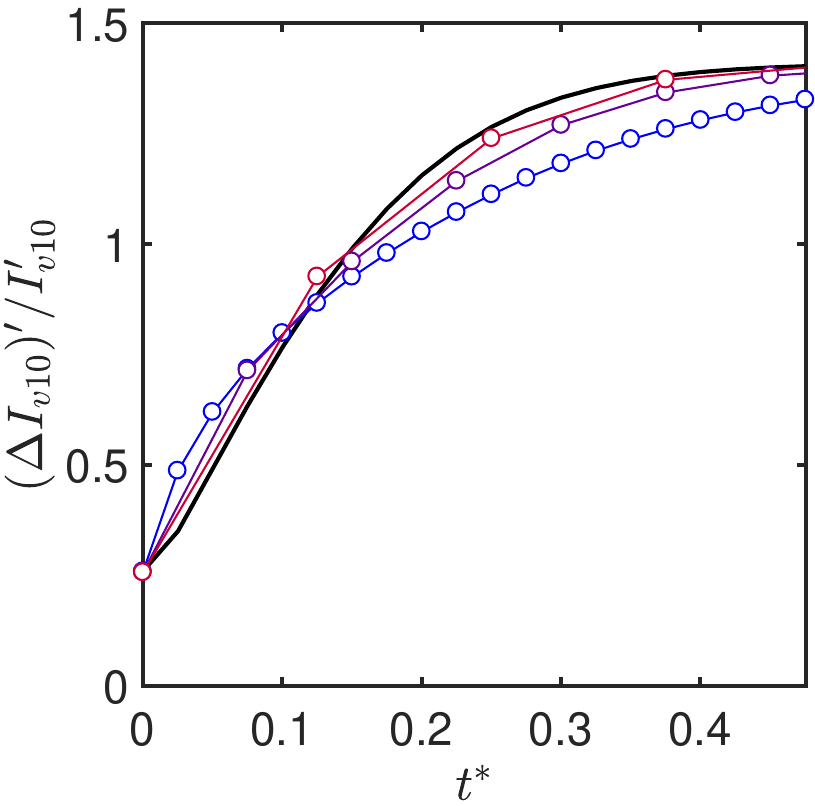}}%
\mylab{-.24\textwidth}{.26\textwidth}{(b)}%
\hspace*{2mm}%
\raisebox{0mm}{\includegraphics[height=.303\textwidth,clip]%
{\figpath 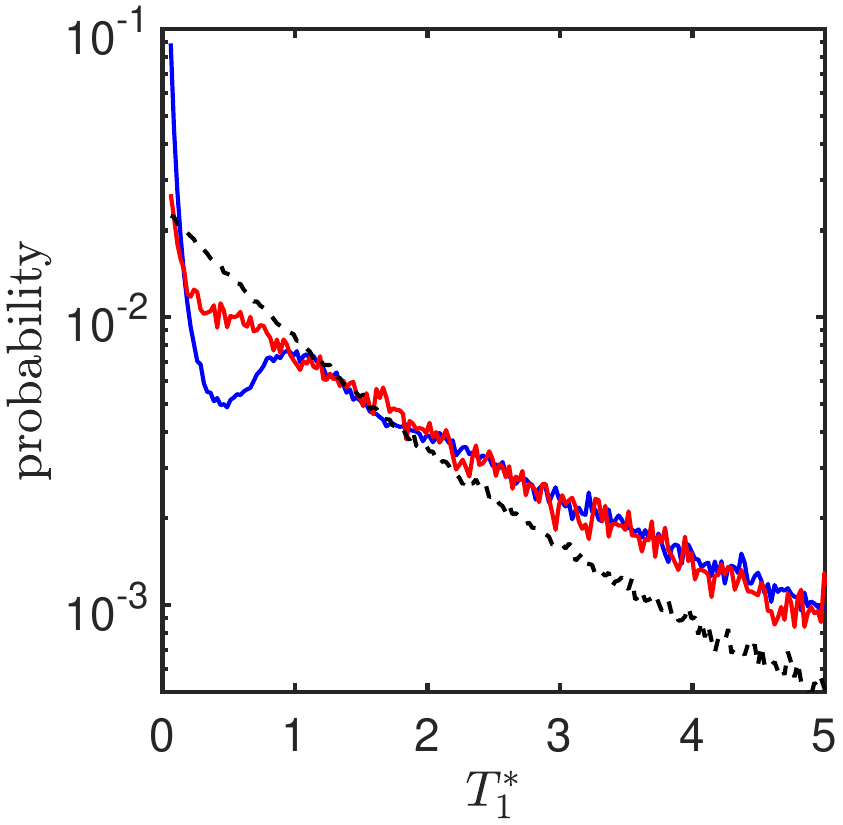}}%
\mylab{-.05\textwidth}{.26\textwidth}{(c)}%
}%
\centerline{%
\raisebox{0mm}{\includegraphics[height=.30\textwidth,clip]%
{\figpath 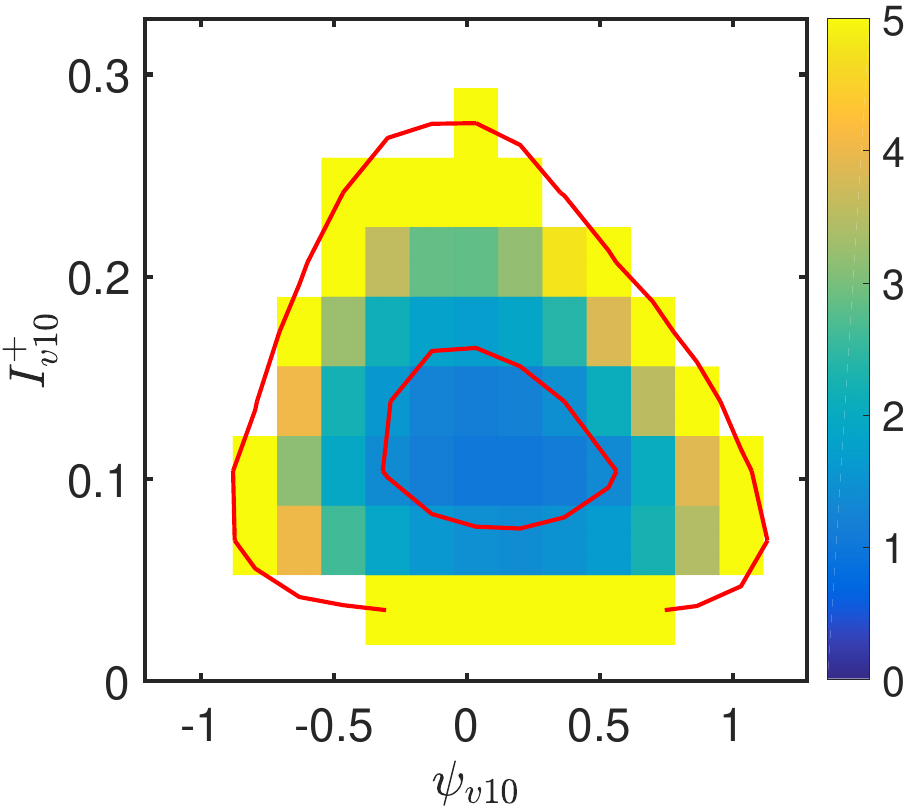}}%
\mylab{-.10\textwidth}{.26\textwidth}{(d)}%
\hspace*{2mm}%
\raisebox{0mm}{\includegraphics[height=.30\textwidth,clip]%
{\figpath 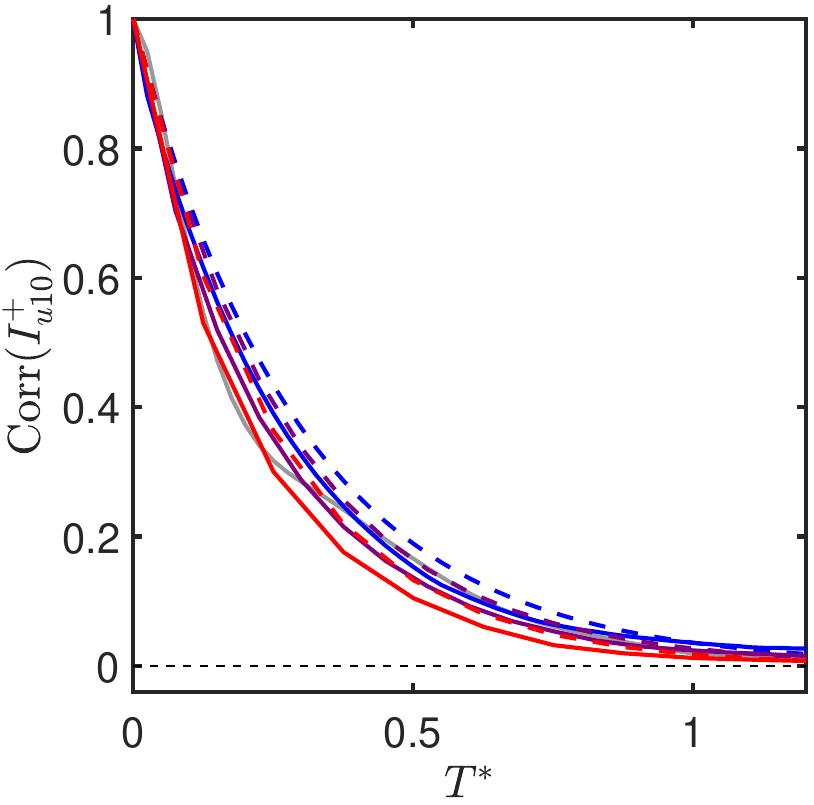}}%
\mylab{-.04\textwidth}{.26\textwidth}{(e)}%
\hspace*{2mm}%
\raisebox{0mm}{\includegraphics[height=.303\textwidth,clip]%
{\figpath 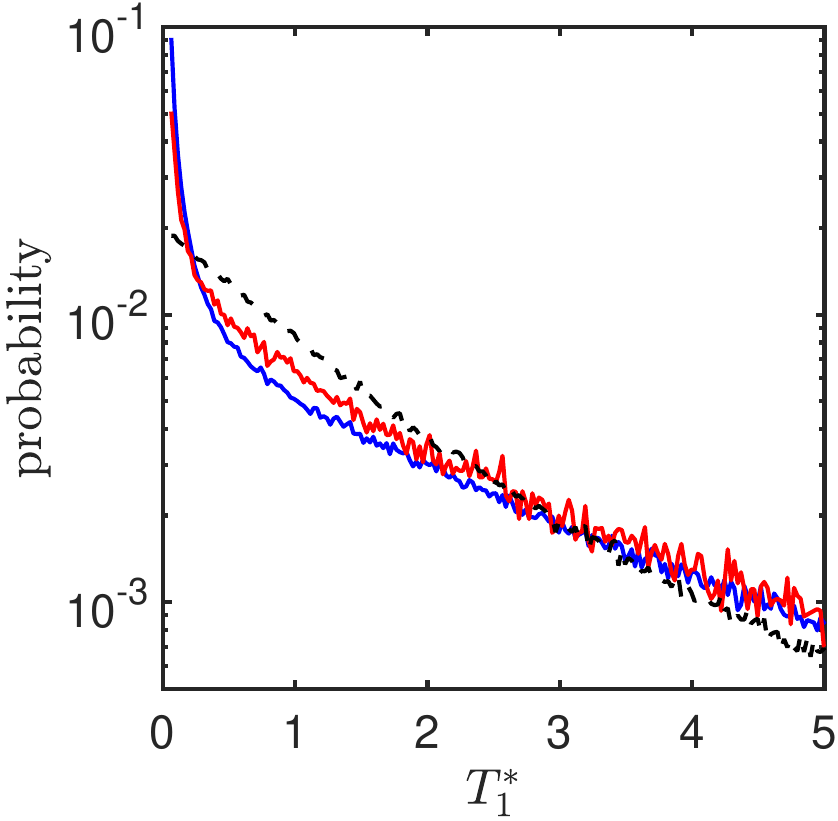}}%
\mylab{-.06\textwidth}{.26\textwidth}{(f)}%
}%
\caption{%
Temporal behaviour of turbulence and of the Markovian model. In (a-d), Case II.
(a) Temporal autocorrelation function of $I_{v01}$. The grey solid 
line is obtained from turbulence. Other solid lines are from the PPF model, and the dashed 
ones are from the subdominant eigenvalue of the PFO. From blue to red: $\Delta t^*=0.025$, 0.075, 0.125. 
(b) Root-mean-squared divergence among trajectories starting from the same partition cell,
normalised with the standard deviation of the variable in question. The solid black line is from the
training data; those with symbols are the PPF model, with colours as in (a).
(c) Probability distribution of the time of first return to individual cells, averaged
within the 95\% probability contour. The continuous blue line is computed from the PPF; the red one is
from the training data, and the dashed one is from a series of cells randomly chosen from the
IPD. For the PPF and random models, $\Delta t^*=0.025$ and $n_T=5\times 10^5$.
(d) Mean return time of the training data for individual cells in (c). 
(e) As in (a), for the disorganised Case I. 
(f) As in (c) for Case I.
In all figures, the partition is $15\times 13$ cells, and C950. 
}
\label{fig:lyap}
\end{figure}

More interesting are the temporal aspects of the flow. Most complex dynamical systems have a range
of temporal scales, from slow ones that describe long-term dynamics, to fast local-in-time events.
In the case of wall-bounded turbulence, a representative slow scale is the bursting period,
$O(h/u_\tau)$ \citep{oscar10_log}. The PFO, which encodes the transition between closely spaced
snapshots, describes the fast time.

Figure \ref{fig:lyap}(a) displays the temporal autocorrelation function, $\bra \vX(t)
\vX(t+T)\ket_t/X'^2$, of one of the model variables, computed independently for the turbulence data and
for the Markov chain of the PPF model. They approximately agree up to $T^*\approx 0.3$.
The correlation of a particular variable depends on how it is distributed over the IPD, but its
decay is bounded by the decorrelation of the probability distribution itself, which approaches the
IPD with the number of iterations as $|\lambda_{\Delta t}|^n$, where $\lambda_{\Delta t}$ is the
eigenvalue of $\bQ_{\Delta t}$ with the second highest modulus \citep{Bremaud}. It is intuitively
clear that, if the distribution of $\bq$ approaches the IPD after a given time interval,
independently of the initial conditions, its correlation with those initial conditions also
vanishes. Figure \ref{fig:lyap}(a) shows that the Markovian models approximately describe turbulence
over times of the order of the probability decorrelation time, which is given by the dashed lines.
The decay of the correlation corresponds to the exponential divergence of nearby initial conditions.
Figure \ref{fig:lyap}(b) shows that the variable in figure \ref{fig:lyap}(a) diverges for
trajectories initially within the same partition cell, averaged over all the cells in the IPD, and
shows that the divergence is complete as the correlation has decays. The PPF model and
its Gaussian approximations reproduce this behaviour reasonably well.

More surprising is that this agreement extends to times of the order of the bursting period, $T^*=O(1)$.
The property of time series that more closely corresponds to periodicity is the
first-recurrence time, $T_1$, after which the system returns to a particular cell. Its
probability distribution is a property of the PFO \citep{feller1XV}, and can be measured from
the time series. Figure \ref{fig:lyap}(c) shows the averaged distribution computed by accumulating
for each cell of the partition the probability of recurring after $T_1$. The red line 
is turbulence data and the blue one is from the PPF model. They agree for very short times, as
expected from figure \ref{fig:lyap}(a,b), and for times longer than a few eddy turnovers. The
dashed black line is a time series in which the order of the cells is randomly selected from
the IPD. The exponential tails of the three distributions suggest that the long-time behaviour of
turbulence and of the PPF is essentially random and memoryless. The discrepancy between the red and
blue lines at  $T_1^*\approx 0.5$ is the same as in the correlations in figure
\ref{fig:lyap}(a), and is characteristic of deterministic projections.
More significant is the probability deficit for $T_1^*\lesssim 2$ between both solid lines and the
randomised dashed line, which is a feature of most variable combinations. That
both turbulence and the PPF preserve this difference shows that the PPF encodes enough information
to approximately reproduce the bursting period, and that bursting, which is responsible for the
longer return periods, is a feature of both turbulence and its PPF approximation. Figure
\ref{fig:lyap}(d) shows the mean return time for individual cells and reveals that long-term
bursting is a property of the periphery of the IPD.

Figures \ref{fig:lyap}(e,f) repeats the analysis in figure \ref{fig:lyap}(a) and \ref{fig:lyap}(c)
for the disorganised Case I. The conclusions from the organised variables also apply to the
disorganised ones, but there are some differences. The dashed lines in figure \ref{fig:lyap}(a) are
the exponential decay due to the subdominant eigenvalue of the PFO. That they depend on the time
interval used in the PFO shows that $\lambda_{\Delta t}^n \ne \lambda_{n\Delta t}$, and the
difference between the two quantities measures the `memory' of the system, which is missing for the
Markovian model. On the other hand, the dashed lines for the three $\Delta t$ in figure
\ref{fig:lyap}(e) essentially collapse, and they also collapse with the decay of the correlation of
the model variable, and even with the turbulence data. This suggests that none of these processes
keeps memory of previous time steps. In fact, while the return plot in figure \ref{fig:lyap}(f)
shows the same probability deficit compared to a random process for short return times as in the
well organised case in figure \ref{fig:lyap}(c), the effect is weaker, and so is the excess
probability in the long tail. This suggests that the time series in Case I are effectively random, in
agreement with figure \ref{fig:maps}(a).

\begin{figure}
\centering
\raisebox{0mm}{\includegraphics[height=.30\textwidth,clip]%
{\figpath 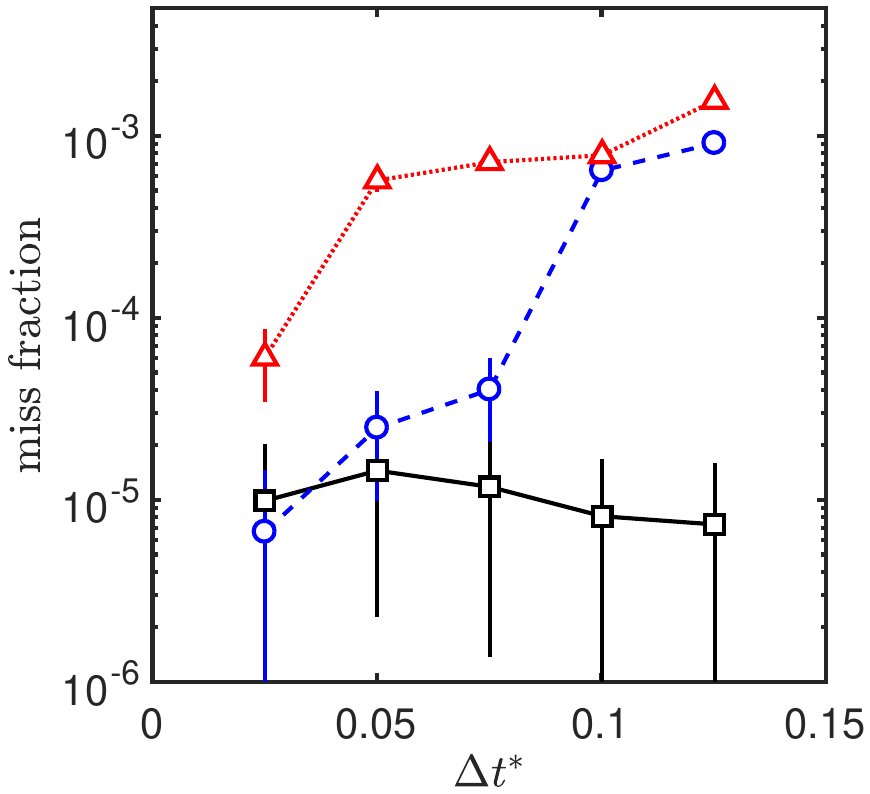}}%
\mylab{-.25\textwidth}{.26\textwidth}{(a)}%
\hspace*{2mm}%
\raisebox{0mm}{\includegraphics[height=.30\textwidth,clip]%
{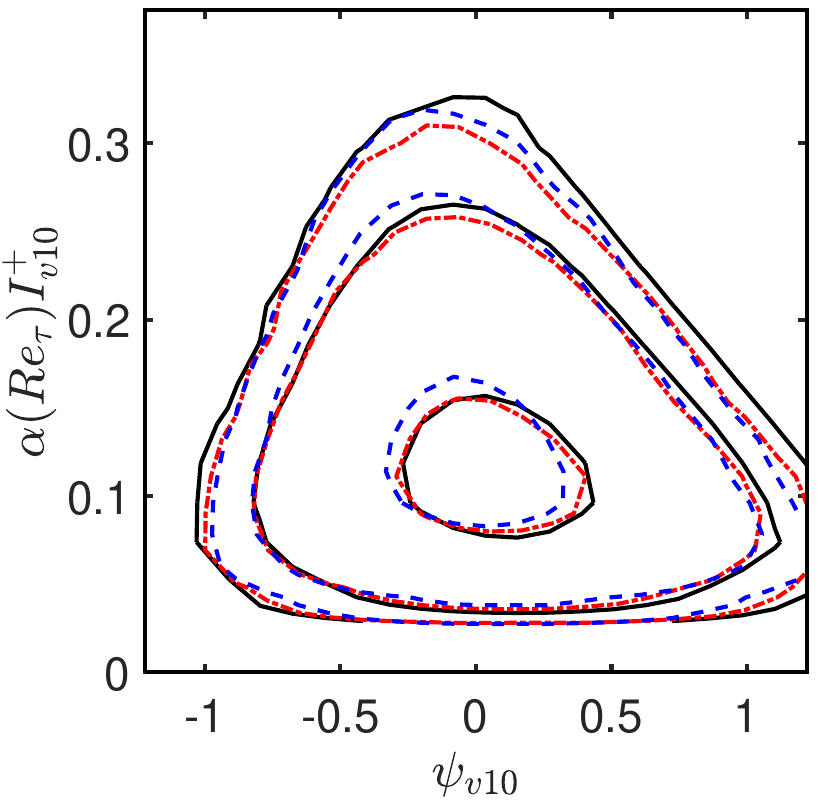}}%
\mylab{-.23\textwidth}{.26\textwidth}{(b)}%
\caption{%
(a) Fraction of required restarts for the three models in figure \ref{fig:fake_PPF}, averaged over 200 experiments. Bars are one standard deviation.
Squares, PPF;  circles, Gaussian approximation to the true PFO;  triangles,
Gaussian approximation with fitted parameters. $n_T=10^5$. $\Delta t^*=0.075$;
$15\times 13$ cells. Case II of C950.
(b) Comparison of the IPDs of Case II for the three Reynolds numbers in table
\ref{tab:cases}. The vertical coordinate is scaled by:
\solid, C950 and $\alpha=1$; \chndot, C550 and $\alpha=0.86$; \dashed, C350 and $\alpha=0.81$. The first and
last stretching factors are manually adjusted for optimum fit. The middle one is linearly
interpolated from the other two.
}
\label{fig:halluc}
\end{figure}

Hallucinations in large language models refer to instances in which they generate plausible but
factually incorrect answers \citep{Rawte:Halluc23}. In the context of our experiment, they happen
when the model drifts into a cell not visited during training, in which case the model is directed
to continue from a randomly selected cell of the IPD. The two upper lines in figure
\ref{fig:halluc}(a) show the fraction of restarts required for the two Gaussian approximations of
the PPF. It grows as the stochastic component of the PFO increases with the time increment, and is
always higher for the globally fitted approximation than for the local one. The lower line is from
the straightforward PPF, trained in a data set that has been broken into shorter sequences to avoid
overfitting. None of the three is large enough to influence the overall statistics.

\section{Conclusions}\la{sec:conc}

We have shown that, at least for quasi-deterministic variable pairs, the one-step PFO acts
as a surrogate for the differential equations of motion and that, in the same way that the latter
generate all the temporal scales of turbulence, the Markov chain induced by the PFO retains
substantial physics over all those scales. We have traced the agreement at very short and very long
times to general properties of Markov chains, but the agreement for times of the
order of an eddy turnover shows that some non-trivial physics is also retained.

Neither the PFO nor its Markov chains can provide information that was not in the original dynamical
system, but they do it more simply. The full PFO is an $N_D^2$ matrix, where $N_D\sim O(200)$ is the
number of cells in the partition. This is already a large reduction from the original number of
degrees of freedom, $O(10^6)$, but the Gaussian approximation is a much shorter list of $5N_D$
numbers, and the quadratic fit to the Gaussian parameters only requires 25 numbers, five for each 
Gaussian moment. This economy, besides simplifying calculations, becomes important when
interpolating models among cases, such as different Reynolds numbers. Although we have mostly
described results for the highest-Reynolds number, C950, most also apply to the
two lower-Reynolds numbers in Table \ref{tab:cases}. An example is figure \ref{fig:halluc}(b), which
compares the invariant densities of the three Reynolds numbers, and shows that they mostly differ by
a rescaling of the intensity axis. This figure also serves as a test for flow interpolation. The
highest and lowest Reynolds numbers in the figure are fitted by hand, but the intermediate one is
linearly interpolated from them as a function of $Re_\tau$. We have finally shown that the PFO can
be substantially modified without much degradation, probably because it is already an approximation.

In essence, the PFO can be understood as a statistical counterpart to the equations of motion, in the
sense that both encode the response of the system at every point in state space. In the case of the
PFO, this is obtained from observation and given in terms of probabilities, while in the case of the
equations it is a functional relation. There are two important differences. The first is that the
PFO works on a submanifold, and cannot make exact predictions. The second, and perhaps most
significant, is that the PFO, derived from passive observations, only has information about the
system attractor, while the equations of motion, which have presumably been
supplemented by experiments outside the attractor, work throughout state space.
In that sense, only the latter would be useful for many control applications. 
 
Perhaps the most intriguing aspect of the discussion above is how little, beyond the initial choice
of a restricted set of variables, is specific to turbulence. Much of the agreement or disagreement
between the models and the original system can be traced to generic properties of the transition
operator, and should therefore apply to other high-dimensional dynamical systems, or to
Markovian models in general.

\vspace{1ex}
{\bf Acknowledgements:}
The author would like to acknowledge informal discussions
with many colleagues over months of perplexity.
This work was supported by the European Research Council under the Caust grant
ERC-AdG-101018287.  
The author reports no conflict of interest.

{\bf Author ORCID:} https://orcid.org/0000-0003-0755-843X

%

\end{document}